\RequirePackage{fix-cm}
\documentclass[runningheads]{llncs2e/llncs}
\usepackage{graphicx,wrapfig}
\usepackage{amssymb}
\usepackage{colortbl}

\usepackage{tikz}
\usepackage{pgfplots}
\pgfplotsset{compat=newest}
\usepgfplotslibrary{units}

\usepackage[linesnumbered,ruled,vlined]{algorithm2e}

\usepackage[linesnumbered,ruled,vlined]{algorithm2e}
\usepackage{caption}
\usepackage{color}
\usepackage{graphicx,wrapfig}
\usepackage{multirow}
\usepackage{pgfplots}
\usepackage[subrefformat=parens,labelformat=parens]{subfig}
\usepackage{tikz}
\usepackage{subfig}
\usepackage{todonotes}
\usepackage{url}
\usepackage{enumitem}

\newcommand{\ignore}[1]{}

\newcommand{\Sch}{\hbox{$\mathbb{G}$}} 			

\newcommand{\source}{\hbox{${\cal S}$}} 
\newcommand{\Q}{\hbox{${\cal Q}$}}
\newcommand{\R}{\hbox{${\cal R}$}}
\newcommand{\C}{\hbox{${\cal C}$}}

\newcommand{\auxR}{\hbox{${L^c_{1,\overline{L_2}}}$}}
\newcommand{\auxRI}{\hbox{${L^c_{2,\overline{L_1}}}$}}

\newcommand{\distDB}{\hbox{$\overline{\source}$}}
\newcommand{\distFDB}{\hbox{$\overline{(\source,\tau)}$}}
\newcommand{\FQ}{\hbox{$q\!:\!\tau_{in}$}}
\newcommand{\FCQ}{\hbox{$q\!:\!\tau_{in}, \C$}}
 
\newcommand{\checking}{\textit{Valid}} 
\newcommand{\cache}{\textit{Cache}}

\def\ie{\textit{i.e.}}
\def\eg{\textit{e.g. }}
\def\wrt{\textit{w.r.t.}}
\def\sqr#1#2{{\vcenter{\vbox{\hrule height.#2pt \hbox{\vrule width.#2pt height#1pt \kern#1pt \vrule width.#2pt}\hrule height.#2pt}}}}

\usepgfplotslibrary{units}
\pgfplotsset{compat=newest}

\definecolor{Gray}{gray}{0.85}
\definecolor{LightCyan}{rgb}{0.88,1,1}

\begin{document}
\title{Querying Linked Data: how to ensure user's quality requirements
}






\author{Jacques Chabin\inst{1} \and
Mirian Halfeld-Ferrari\inst{1} \and
B\'{e}atrice Markhoff\inst{2} \and
Thanh Binh Nguyen\inst{3}}
\authorrunning{Chabin, Halfeld-Ferrari,Markhoff, guyen}
\institute{Universit\'e Orl\'eans, INSA CVL, LIFO EA, Orl\'eans, France
\and
 Universit\'{e} de Tours, LIFAT, Blois, France.
 \and
	Quy Nhon University, Vietnam.\\
\email{\{jchabin,mirian\}@univ-orleans.fr, beatrice.markhoff@univ-tours.fr nguyenbinh@qnu.edu.vn}}

\maketitle

\begin{center}
\today
\end{center}

\begin{abstract}
In the distributed and dynamic framework of the Web, data quality is a big challenge. The Linked Open Data (LOD) provides an enormous amount of data, the quality of which is difficult to control. Quality is intrinsically a matter of usage, so consumers need ways to specify quality rules that make sense for their use, in order to get only data conforming to these rules. We propose a user-side query framework equipped with a checker of constraints and confidence levels on data resulting from LOD providers’ query evaluations. We detail its theoretical foundations and we provide experimental results showing that the check additional cost is reasonable and that integrating the constraints in the queries further improves it significantly. 
\keywords{Query rewriting \and Constraints \and Context \and Confidence degree \and User-side~quality}
\end{abstract}

\section{Introduction}
\label{intro}

Very large knowledge bases on the  web of Linked Open Data (as DBpedia, Yago or BabelNet) need 
applications to  help humans  exploring their huge knowledge,   performing data analysis and data mining tasks.
One crucial point for such applications, and in particular for data mining algorithms, is that the data collection and pre-processing steps have to be \textit{safe and sound}.


In order to help semantic Web data mining tool designers for performing the data collection and pre-processing steps, we propose a semantic web data validator. 
The idea is to extend a query environment over semantic graph databases with a mechanism for filtering answers according to a user customized context. 
The \textit{user context} is composed of \textit{(i)} the \textit{view} she/he has defined on the needed semantic web data and \textit{(ii)} a set of \textit{personalization tools}, such as integrity constraints and confidence degrees.
Both the constraints and the queries are expressed in terms of the user's view of data.
Query answers are built only with data respecting a required confidence degree.
 The constraint verification is triggered by a query and consists in filtering its answers. 
 In this way, there may be some inconsistencies within sources, but the answers given to the user are filtered to ensure their consistency \wrt\  her/his context.

\paragraph{Motivating example.}

Let \textsf{Univ} be a database where :
$professor(X_{id})$ and $employ-$ $eeGov(X_{id})$ indicate if a person is a professor or an employee working for the government; 
$teacherOf(X_{id},X_{course})$ associates a teacher with a course,
while $takesCourse(X_{id},X_{course})$ refers to a student enrolling in a course;
$offered-$ $CourseAt(X_{course},X_{dep})$ indicates in which department a course is taught; 
$re-$ $searchesIn(X_{id},X_{domain})$ and $worksFor(X_{id}, X_{dep}, X_{org})$ refer to the research domain and the work place (a department of an organization) of a person;
$headOf (X_{id}, Y_{dep})$ indicates the head of a department.
The context depicted in  Table~\ref{tabConst} is composed by $6$ constraints ($4$ positive, $1$ negative and $1$ EGD), namely:
($c_{P_1}$) every professor must offer a course;
 ($c_{P_2}$) every course given by a teacher must be associated with a department;
($c_{P_3}$) professors are government employees;
 ($c_{P_4}$) if a teacher offers a database course, then he must be a researcher in the database domain;
 ($c_{N_1}$) nobody can teach and follow the same course;
 ($c_{K_1}$) the head of a department in an organization cannot be a person working in a different department.
\vspace*{-0.3cm}

\begin{table}[hb]
	\centering
	\footnotesize
	\begin{tabular}{ll} \hline
		$\C_P$&\textsc{positive constraints}\\ \hline
		$c_{P_1}:$ & $professor(X_{id}) \rightarrow teacherOf (X_{id},X_{course}).$\\
		$c_{P_2}:$ & $teacherOf(X_{id},X_{course}) \rightarrow offeredCourseAt(X_{course},X_{dep}).$\\
		$c_{P_3}:$ & $professor(X_{id}) \rightarrow employeeGov(X_{id}).$\\ 
		$c_{P_4}:$ & $teacherOf(X_{id},DB) \rightarrow researchesIn(X_{id},DB).$\\ 
		 \hline
		
		$\C_N$&\textsc{negative constraints}\\ \hline
		$c_{N_1}:$ &$ teacherOf(X_{id},X_{course}), takesCourse(X_{id},X_{course}) \rightarrow \bot $ \\ 
		\hline
		
		$\C_K$&\textsc{key constraints}\\ \hline
		$c_{K_1}:$&$ worksFor(X_{id}, X_{dep}, X_{org}), headOf (X_{id}, Y_{dep}) \rightarrow X_{dep}=Y_{dep}$ \\ 
	\end{tabular}
	\normalsize
	\caption{Set of constraints on \Sch \label{tabConst}}
\end{table}

\vspace*{-0.5cm}
The querying system computes answers on a distributed database, where data sources come from different providers who are  not trusted equally.
In Table~\ref{figEx1},   $\tau$ indicates the accuracy associated with each provider
(Source 1 is considered to be $95\%$ reliable while the reliance on Source 3 is smaller).

\begin{table}[h]
	\tiny
	\centering
	\begin{tabular}{|l|c|l|c|l|}
		\cline{1-1} \cline{3-3}	\cline{5-5}
		{\bf Source 1}, {\bf $\tau_{S1}=0.95$} &\hspace{0.3cm} & {\bf Source 2}, {\bf $\tau_{S2}=0.80$} &\hspace{0.3cm} & {\bf Source 3}, {\bf $\tau_{S3}=0.70$} \\ \cline{1-1} \cline{3-3}	\cline{5-5}
		professor(Bob)			&\hspace{0.3cm}&	offeredCourseAt(DB,LIFO)&\hspace{0.3cm}&	professor(Peter)	\\
		professor(Tom)			&\hspace{0.3cm}&	offeredCourseAt(Java,LIFAT)&\hspace{0.3cm}&	professor(Ann)		\\
		professor(Alice)		&\hspace{0.3cm}&	worksFor(Bob,LIFO,UO)	&\hspace{0.3cm}&	headOf(Bob,LIFO)	\\
		bornIn(Bob,USA)	&\hspace{0.3cm}&	worksFor(Ann,LIFAT,UT)		&\hspace{0.3cm}&	headOf(Ann,CNRS)	\\
		bornIn(Tom,UK)	&\hspace{0.3cm}&	takesCourse(Tom, Java)	&\hspace{0.3cm}&	bornIn(Peter,UK)	\\
		bornIn(Alice,Denmark)	&\hspace{0.3cm}&	takesCourse(Bob, Net)	&\hspace{0.3cm}&	bornIn(Ann,USA)	\\
		foreignCountry(USA)		&\hspace{0.3cm}&	teacherOf(Bob, DB)		&\hspace{0.3cm}&	teacherOf(Peter,Java)	\\
		foreignCountry(UK)		&\hspace{0.3cm}&	teacherOf(Bob, Java)	&\hspace{0.3cm}&	teacherOf(Ann,DB)	\\
		foreignCountry(Denmark)	&\hspace{0.3cm}&	teacherOf(Tom, Java)	&\hspace{0.3cm}&	researchesIn(Bob,DB)	\\
								&\hspace{0.3cm}&	teacherOf(Alice, Net)	&\hspace{0.3cm}&	employeeGov(Bob)	\\
								&\hspace{0.3cm}&							&\hspace{0.3cm}&	employeeGov(Tom)	\\
								&\hspace{0.3cm}&							&\hspace{0.3cm}&	employeeGov(Alice) 	\\
								&\hspace{0.3cm}&							&\hspace{0.3cm}&	employeeGov(Peter)	\\
		\cline{1-1} \cline{3-3}	\cline{5-5}
	\end{tabular}
	\caption{Example of local sources \label{figEx1}}
\end{table}
\normalsize

Suppose one user wants to work in a context that contains only constraints $c_{P1}$ and $c_{P2}$. \ignore{and $c_{N1}$}
In this scenario let us consider query $q(X) \leftarrow professor(X), $ $bornIn(X,Y), foreignCountry(Y)$
to find the foreign professors. 
The required confidence degree is $\tau_{in}= 0.75$, indicating that sources having a smaller confidence degree should not be taken into account.
In this case, only data from sources 1 and  2 are considered. Source 3 is ignored as $\tau_{S3} < \tau_{in}$. 
The answer is the set $\{(Bob):0.8, (Tom):0.8\}$ as both Bob and Tom are professor ($professor(Bob), professor(Tom)$); 
they were born in foreign countries ($bornIn(Bob,USA)$, $bornIn(Tom,UK)$, $foreignCountry(USA)$, $foreign\-$ \linebreak$Country(UK)$), 
and they teach at least one course ($teacherOf(Bob, DB)$,\linebreak $teacherOf(Tom, Java)$) at a department 
($offeredCourseAt(DB,LIFO)$, \linebreak $offeredCourseAt(Java,LIFAT)$). 
Tuple 
$(Alice)$ is not an answer because, although  we have  $teacherOf(Alice, Net)$ in the database ($c_{P1}$ is satisfied), her course is not associated with any department, a violation of constraint $c_{P2}$. 

A new user wants a context  involving  all the constraints in Table~\ref{tabConst}. 
For  a degree $\tau_{in}= 0.7$ and the  same query of the first scenario, data from all three sources are taken into account.
$Alice$'s course, $Net$,  is  associated with no department (violating $c_{P2}$), $Tom$ teaches  and follows a $Java$ course (violating $c_{N1}$),  $Ann$ is not a government employee (violating $c_{P3}$). 
In fact, $c_{P4}$ and $c_{K1}$ are also violated as \textit{Ann} teaches database but she does not do research in the database domain. Moreover, she is the head of $CNRS$ while not working there. 
Only $Bob$ and $Peter$ satisfy all constraints. Thus, the answer is the set $\{(Bob):0.7, (Peter):0.7\}$.~$\hfill\Box$

\paragraph{Querying Environment.}
Our query processing framework, depicted in Figure~\ref{figOverview},  comprises two distinct parts which communicate: \textit{Data validation}, responsible for checking constraints satisfaction, and \textit{Data providers} for computing answers to the queries. The latter may integrate several end-data-providers, or it may connect only one provider.
For ensuring that the final answers to the user's queries satisfy all user constraints, a dialogue between the two parts is established, for getting intermediate results and sending subsidiary queries.  

A user  \textit{context} sets the user's view on the queried sources. It  is defined as  set of predicates and 
quality constraints involving these predicates.
The user's \textit{query} also involves these predicates,  so
quality constraints can be used as rewriting-rules to reformulate each query $q$ into a set of conjunctive queries whose answers, contained in $q$'s answers, are valid w.r.t. the user quality constraints. 
Afterwards, these conjunctive queries are sent to the \textit{Data providers} part, which evaluates them against data stored  on sources. 
The  query evaluation process is transparent to the validation step, in particular, answers that are entailed are treated in the same way as those that actually exist in sources. We respect the potential ontological dimension of semantic web sources, while interpreting the \textit{user constraints} using the closed-world assumption. 
Indeed, as it deals with semantic data, the evaluating process performed by the \textit{Data providers} part relies on the open-world assumption, where ontological constraints are used to deduce new information.
Ontological constraints are used as rewriting-rules to reformulate a query into a set of new conjunctive queries, 
for taking into account integration information (OBDA/OBDI Systems \cite{poggi2008linking,Abiteboul11}), or for dealing with incomplete information issues \cite{Abiteboul11,GOP11,Lembo20153}. But such rewritings are performed by the \textit{Data providers} part, independently from the \textit{Data validation} part.  
As our system may be deployed with various data management systems, a translator module deals with datalog+- \cite{CGL12} for Graal\footnote{\url{https://graphik-team.github.io/graal/}}, SPARQL for FedX \cite{Schwarte11}, and HIVE-SQL for MapReduce (see~\cite{BinhTHESE}). 

\noindent
\textit{Organization (Extended Version).} This paper extends the work in~\cite{Binh2018} by presenting a new and more detailed version of our rewriting querying approach together with its theoretical foundations and new experimental results.
It is organised as follows: Section~\ref{relWork} focus on related work.
Section~\ref{section:ValidAnswerFormalDefinition} formally introduces our querying system.
Section~\ref{section:RewritingApproach} details our query rewriting system and Section~\ref{chapter:Experiments} presents experimental results.
Section~\ref{sec:conclusions} concludes the paper.


\section{Related Work}
\label{relWork}
\noindent
The main goal of this work is to provide a quality-driven querying system on distributed data, particularly on LOD. But our proposal is independent of LOD specificities, as the results verification is based on a dialogue between the checker and a LOD endpoint (or a federated query system).  
%
Inconsistency in knowledge bases is the kernel of proposals such as~\cite{Cali2003,CHL20,Goasdoue13,Lembo05,Lukasiewicz2012,Rosati2012}, which consider the \textit{consistency of the data stored in the source databases}.
The  general idea is to \textit{restore} database consistency or to propose \textit{update} approaches which ensure consistency or to \textit{refuse} the database when consistency is not achievable. 
In brief, those proposals try to use constraints to compute the new consistent database.
Contrary to them, our work focuses on the \textit{consistency of the query answers returned to a user}. 
	Instead of trying to establish the consistency of a database in order to query over it, 
	we want to, firstly, just evaluate a query on a given (possibly) inconsistent database, 
	and, then, use rules (which define a context) as filters to obtain the valid answers. 
	In this way, we also obtain \textit{consistent data}.
	Our goal is neither to modify the database nor to try to compute its different possible consistent states. 
	We do not want to refuse an inconsistent database. 
	Since the meaning of consistency is in connexion to a user profile; 
		one can admit things that are prohibited by others. 
	Checking consistency of the answers is usually a smaller task than verifying the whole database validity.
	A parallel can be done between saturation in \cite{Goasdoue13}, the chase in \cite{CHL20} and our query rewriting.

\begin{figure}
\tiny
\centering
\begin{tabular}{|m{5cm}|m{6cm}|}\hline
\subfloat[Query system overview.]{\label{figOverview}\includegraphics[scale=0.4]{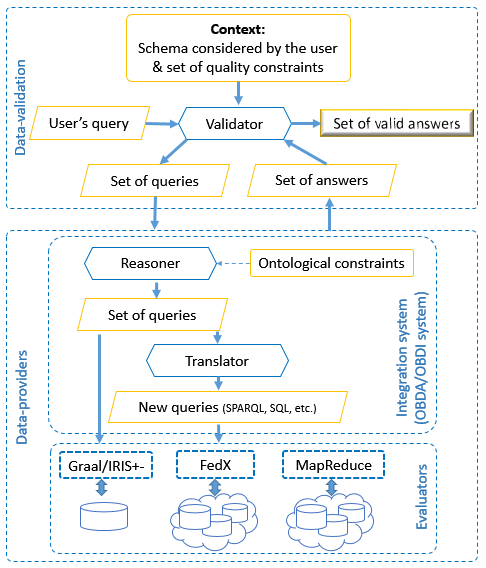}} &
\subfloat[Module Validator.]{ \label{figDetail}\includegraphics[scale=0.2]{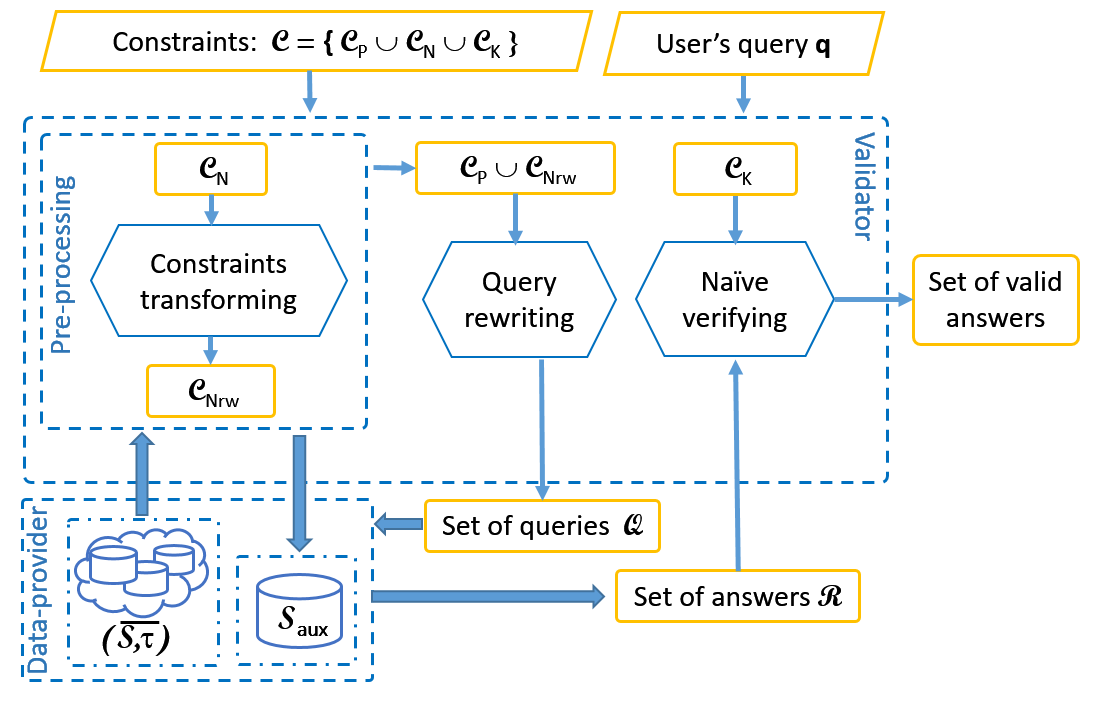}}\\ \hline
\end{tabular}
\caption{Query system overview and a zoom-in the Validator module.}
\label{Fig1}
\vspace*{-0.5cm}
\end{figure}

Besides, in our proposal, rules defining a context are quality constraints. 
They impose a filter on retrieved data (in contrast to work, \eg~\cite{Cali2003,Lukasiewicz2012,Rosati2012}, 
 where constraints are just inference rules). 
In \cite{Deutsch06} we find  a solution that is close to ours, but
it does not comprehensively consider the presence of constants in constraints and queries and 
it does not handle negative constraints, either.
Moreover, the way it deals with equality-generating dependencies (EGD) differs from ours.
Precisely, in~\cite{Deutsch06} an equality is added to the rewritten query only when the whole body of an EGD matches a subset of atoms in the query's body. This is an EGD chase step~\cite{Onet13}. 
In our solution, an EGD has to be considered even if there is only one atom in EGD's body matching an atom in the query's body, reinforcing the idea that
only consistent data are allowed to compute results.
Finally, the principle of restricting results by rules is also the motivation of \cite{Cima2020}, which extends ontology-based data access (OBDA) with data protection policies. But they proceed by compiling the policies into the OBDA mappings, whereas our proposal is independent of whether or not the query evaluation uses the ontological part of LOD resources (see Fig.\ref{Fig1}).


\section{Querying System: formal definitions}
\label{section:ValidAnswerFormalDefinition}

Our constraints are expressed in a first-order logic formalism.
We consider an alphabet  made up of the equality symbol ($=$), quantifiers ($\forall$ and $\exists$),  the symbols $\top$ (true) and $\bot$ (false) and of 
four disjoint sets 
\textsf{const} (of constants)
 \textsf{null} (of fresh labeled \textit{nulls}),
\textsf{var} (of variables  used to range over elements of \textsf{const} or \textsf{null})
and \textsf{pred} (of predicates).
The only possible \textit{terms} are constants, nulls or variables.
A \textit{free tuple} $u$ is a sequence of either variables or constants, or both. 
We denote by $var(u)$ the set of variables in $u$.
An \textit{atom} is a formula having  one of the forms:
(i) $p(t_1, . . . , t_n)$, where $p$ is an $n$-ary predicate, and $ t_1 , . . . , t_n$ are terms;
(ii)  expressions $\top$ and $\bot$; 
(iii)  $t_1 = t_2$, where $t_1$ and $t_2$ are terms.
A conjunction of atoms is often identified with the set of all its atoms.
A fact over a predicate $p$ is an atom $p(u)$ where $u \in \textsf{const}^n$. 
A {\em homomorphism} from a set of atoms $A_1$ to a set of atoms $A_2$ is a mapping $h$ from the  terms of $A_1$ to the terms of $A_2$ such that:
(i)  if $t \in \textsf{const}$, $\bot$ or $\top$  then $h(t) = t$;
(ii)  if $t \in \textsf{null} $, then $ h(t) \in (\textsf{const} \cup \textsf{null} )$ and
(iii) if $r(t_1,...,t_n)$ is in $A_1$, then $h(r(t_1,...,t_n)) = r(h(t_1), . . . , h(t_n))$ is in $A_2$.
The notion of  homomorphism naturally extends to sets of atoms and conjunctions of atoms.
An \textit{isomorphism} is a bijective homomorphism.
Two atoms $A$ and $A'$ are \textit{unifiable} iff  there exists an homomorphism $\sigma$ (denoted by \textit{unifier}) on $\{A,A'\}$,   such that  $\sigma (A) = \sigma(A')$. 
If $A$ and $A'$ are unifiable, then they have a unifier $\theta$ such that every unifier $\sigma$ of $A$ and $A'$ can be written as $\sigma = h \theta$ for some homomorphism $h$ on $\{A,A'\}$.  Such a unifier is called the \textit{most general unifier} of $A$ and $A'$ (denote by \textit{mgu}).

A \textit{conjunctive query} (CQ)    has the form
$q: R_0 (u_0) \leftarrow R_1(u_1) \dots  R_n(u_n) \linebreak comp_1(v_1) \dots comp_m(v_m)$
	where 
	$n,m \geq 0$, $R_i$ ($0 \leq i \leq n$) are predicate names of a database schema,  
	$u_i$ are  free tuple 
	of 	appropriate arity 
	and $comp_j(v_j)$  ($0 \leq j \leq m$) are comparison formulas having the form $(X=a)$ or $(X=Y)$ where $X$ and $Y$ are variables appearing in $u_i$ ($0 \leq i \leq n$) and $a$ is a constant.  
We denote $head(q)$ its  left hand-side  and  $body(q)$ its   right hand-side.
Queries are \textit{range restricted} and  \textit{satisfiable}~\cite{AHV95}. 
The \textit{answer} to a conjunctive query $q$ of arity $n$ over a database instance $I$, denoted as $q(I)$, is the set of all tuples $t \in \textsf{const}^n$ for which there exists a homomorphism 
$h_t$ such that $ h_t(body(q)) \subseteq I$ and $h_t(u_0) = t$.\\
(*) We denote by $h_t$ \textit{a homomorphism used to obtain an answer tuple}~$t$.

\vspace*{-0.3cm}
\paragraph{Querying with confidence.}
Confidence in data providers may vary and a user's query $q$ may be bounded by a confidence degree $\tau_{in}$.
We define answering with confidence on a data source and then on a set of data sources, possibly from different providers.
Indeed, $q$'s evaluation may concern distinct data sources.

\begin{definition}[Querying with confidence on a data source]
	\label{locaCD}{\rm
		Let $(\source,$ $ \tau)$ be a source database: $\source$ is a database instance and $\tau$ is its truth or confidence degree. 
		Let $q$ be a query  and  $\tau_{in}$ its truth degree.
		The answer of $q:\tau_{in}$ over $(\source, \tau)$ is the set 
		$ans(\FQ, (\source, $ $\tau)) =$ $\{ (t:\tau_{out}) \mid \tau_{out}=\tau$ and $t \in q(\source)$ and $cond(\tau_{in},\tau)\}$,
		where $cond(\tau_{in},\tau)$ is a condition given by the user.
		$\hfill\Box$
	}
\end{definition}


\begin{definition}[Querying with confidence on  $\distFDB$ ]{\rm
		\label{def-CandAns}
		Let $\distFDB$ be a  database instance composed of $n$  databases having different truth degrees.
		A couple $(t:\tau_{out})$ is a \textit{candidate answer} for  query 
		($q:\tau_{in}, \distFDB)$ if the following conditions hold:
		$(1)$ tuple $t$ is an answer obtained from local sources, \ie\ $t \in ans(q, \distDB)$;
		$(2)$ $\tau_{in} \leq \tau_{out}$ and 
		the computation of $\tau_{out}$ is defined by 
		$\tau_{out} = f(\tau_{in}, \{ \tau^1_{out_{S_i}}, \dots, \tau^m_{out_{S_l}}\})$
		where: 
		$(i)$ each $\tau^k_{out_{S_j}}$ denotes the degree of the tuples in 
		$ans( q_k:\tau^k_{in}, (S_j, \tau_j))$ for the sub-query
		$q_k$ ($1 \leq k \leq m$) generated to be evaluated on the local source $(S_j, \tau_j)$
		during the evaluation process of $q$ (where $i, j, l \in [1,n]$) and
		$(ii)$ function $f$ computes a confidence degree taking as input the query confidence degree 
		and the confidence degrees of data sources concerned by $q$.~$\hfill\Box$
		
	}
\end{definition}

A user can parametrize the use of confidence degrees by choosing different functions $f$ and by combining this choice with $cond$
in Definition~\ref{locaCD}. 
In this paper we  consider that $f(\tau_{in}, \{ \tau^1_{out_{S_i}}, \dots, \tau^m_{out_{S_l}} \})$ corresponds to $min (\{ \tau^1_{out_{S_i}},\dots,$ $ \tau^m_{out_{S_l}}\})$
and  we disregard sources whose confidence is inferior to $\tau_{in}$.

\paragraph{Context-driven querying system.}
A  context is  a set of constraints.
Only data respecting them are allowed as  answers for the user's query.

\begin{definition}[Constraints]
\rm
\label{defConstraint} A  \textit{context}  \C\   is composed by the following sets:

	\noindent
	$\bullet$
	\textit{\textbf{Positive constraints}} ($\C_P$): A positive constraint has the form \\
	\centerline{$c: \forall x_1, \dots, x_n ~~~ L_1(u_1) \rightarrow \exists y_1, \dots, y_m ~~ L_2(u_2)$}
	where 
	$L_1 $ and $ L_2$ are predicate names in $\Sch$;
	$u_1$ and $u_2$ are free tuples  and  
	$var(u_1) \cap var (u_2) = \{ x_1, \dots, x_k \}$  for
	 	$var(u_1) = \{ x_1, \dots, x_k, x_{k+1}, \dots, x_n \} $, 
		\linebreak			$var(u_2) = \{ x_1, \dots, x_k, y_1, \dots, y_m \} $, with $n \geq k \geq 0$, $m \geq 0$.

	\noindent
	$\bullet$ 
	\textit{\textbf{Negative constraints}} ($\C_N$): A negative constraint is a rule where all variables  in the free tuple $u$ 
	are universally quantified and which has one of the (equivalent) formats: 
$c: \phi(u)  \rightarrow \bot$ or  $c: \phi'(u), comp \rightarrow \bot$.
	Formula $\phi(u)$ is an atom $L_1(u)$ or a conjunction of two atoms $L_1(u_1), L_2(u_2)$, for which
	if $var(u_1)$ and $var(u_2)$ are non-empty sets, then $var(u_1) \cap var(u_2) \neq \emptyset$.
	In the second format,  \textit{equalities} on  variables are explicitly expressed in the \textit{comp} atoms.
	We refer to $\C_{N1}$ and $\C_{N2}$ as sets of negative constraints having only one atom and two atoms, respectively.

	\noindent
	$\bullet$ 
	\textit{\textbf{Equality-generating dependency constraints}} without nulls ($\C_K$) (also called key constraints): An EGD is a rule having the general form \linebreak
$c: ~~ L_1(u_1), L_2(u_2) \rightarrow u'_1 = u'_2 $ 
	where 
	all variables in the free tuple $u_1$ and $u_2$ are universally quantified; 
	$L_1 $, $ L_2$ are predicate names in $\Sch$;
	$u_1$, $u_2$ are free tuples such that $var(u_1) \cap var(u_2) \neq \emptyset$;
	$u'_1$ and $u'_2$ are sub-tuples (\ie\ an ordered-subset of variables) of var($u_1$) and var($u_2 $) respectively. 
	EGD include functional dependency (and thus, key constraints) having the form: 
	$L_1(u_1), L_1(u_2) \rightarrow u'_1 = u'_2 $ $\hfill\Box$
\end{definition}

For simplicity, we usually omit quantifiers.
We say that a constraint $c$ is triggered by an atom $A$ when there is a homomorphism $h$ from $body(c)$ to $A$. Positive constraints are a special case of  linear tuple  generating dependency (TGD \cite{AHV95})   which contain only one atom in the head. When $u_2$ has existential quantified variables, the homomorphism $h$ is extended to $h'$ such that, for each existential variable $z \in u_2$, $h'(z) $ is a new fresh null. It is well-known that facts from a database instance may trigger such constraints, and  
the \textit{chase} procedure \cite{Maier79}  is the standard process for the generation of new facts from a database instance and a set of dependencies (TGD or EGD) \cite{AHV95}. 
It can also be used to decide  containment of conjunctive queries in the presence of constraints \cite{JK84}.
 We consider that the set of positive constraints is weakly acyclic~\cite{Fagin2005}, ensuring:
 (i)   that  positive constraints \textit{do not have a cyclic condition such that a null value forces the adding of a new null value}~\cite{Onet13} and 
 (ii)  the decidability  of query containment \cite{Deutsch06}.
 Our motivating example in Section~\ref{intro} illustrates all three types of constraints.

Now, let \FQ\ be a CQ and $\C = \C_P \cup \C_N \cup \C_K$ be a context. 
Valid  answers are those that respect constraints in \C\ and are obtained by trusted databases.
In the following, given $\C_P$, we denote by $chase(\C_P, I)$ the procedure capable of computing all consequences of $\C_P$\ in an instance $I$.   We refer to ~\cite{Benedikt17,Onet13} for details on the different chase algorithms. 

\begin{definition}[Valid  answers]
	\label{def:ValAns}
	{\rm
		The set of valid  answers of a query $q:\tau_{in}$, restrained by \C, over a database \distFDB,
		denoted by $valAns$ $(\FCQ,$ $\distFDB) $ is defined by the set 
		$ \{ (t:\tau_{out}) \}$ respecting the following conditions:\\
	$\bullet$	 $t$ is a candidate answer as in Definition~\ref{def-CandAns} with its (maybe several) $h_t$~(*); \\
$\bullet$ there exists $h_t$ such that for all $L \in$ $h_t( chase(\C_P,J))$, where $J=\{ l | l \in h_t(body(q)) \}$, the following conditions hold:\\
(a) there is a positive answer for $q() \leftarrow L: \tau_{in}$ on $\distFDB$; \\
(b)  for each $c \in \C_{N2}$ of the form $L_1, L_2 \rightarrow \bot $,
				if there is a homomorphism $\nu$ such that 
				$\nu(L_i) = L$, then there is no homomorphism $\nu'$ that extends $\nu$ and for which there is a positive answer for 
				$q'() \leftarrow \nu'(L_{\overline{i}}): \tau_{in}$ (In our notation, if $i= 1$ then $\overline{i} = 2$ and vice-versa); \\
(c)  for each $c \in \C_{N1}$ of the form $L_1 \rightarrow \bot $,
				there is no homomorphism $\nu$ such that $\nu(L_1) = L$;\\
(d)  for each $c \in \C_{K}$ of the form $L_1(u_1), L_2(u_2) \rightarrow X_1 = X_2$ where $X_1$ and $X_2$ are variable in $u_1$ and $u_2$ respectively,
				if there is a homomorphism $\nu$ such that $\nu(L_i(u_i)) = L$, then the answer of 
				$q(X_{\overline{i}}) \leftarrow \nu(L_{\overline{i}}(u_{\overline{i}})): \tau_{in} $
				is a singleton containing the tuple value $\nu(X_i)$. $\hfill\Box$

	}
\end{definition}

Valid answers can be computed in different ways. 
The \textbf{naive approach} is a direct implementation of Definition~\ref{def:ValAns}: its  main idea   is to validate each answer of a query $q$ by generating sub-queries corresponding to each constraint $c \in \C$. 
We refer to~\cite{BinhTHESE} for the  algorithm implementing this approach. 
The rewriting approach focuses on building new queries that incorporate the constraints. It is the subject of the following section.


\section{The query rewriting approach}
\label{section:RewritingApproach}

Figure~\ref{figDetail} offers a closer look at the \textit{Validator}  module.
It  receives the user's query $q$ and a set  \C\ of  constraints, and treats each constraint type differently. 
In a preprocessing step, the set of negative constraints $\C_N$ is translated into a  new set $\C_{Nrw}$ of positive constraints  involving new  auxiliary relations. 
Auxiliary relations are stored in $S_{aux}$, an \textit{auxiliary database}, or  they can  be defined as views and evaluated over the database instance when needed. 
Translated negative constraints, together with positive constraints, are used as rewriting rules:
if the body of a constraint $c$ matches atoms in the body of  $q$, then the head of $c$ is integrated into $q$'s body.
The new rewritten query allows the \textit{validation} and the \textit{query evaluation} to be  performed together,  avoiding validating an extremely large number of intermediate results.
This query rewriting process  gives rise to a set  $\Q$ of rewritten queries, which are evaluated  over  $S_{aux}$ and \distFDB.
If we denote by $\R$ the set of answers obtained from the evaluation of $\Q$, then $\R$ is valid  \wrt\  $\C_P$ and $\C_N$ (proofs in Section~\ref{subsection:Proof}). 
The validity of each  tuple in $\R$  \wrt\ key constraints is done by the naive approach.

Before presenting our rewrite method in details, we consider three examples to illustrate the situations our  algorithm tackles with. 

\begin{example}
	\rm
	\label{ex_case1} 
	Let \small $q_1(X_1) \leftarrow professor(X_1), bornIn(X_1, Z_1), foreignCoun$\-$try(Z_1)$ \normalsize be the query considered in Section~\ref{intro} and a user's context  composed of $c_{P_1}$, $c_{P_2}$ and $c_{P_3}$ (Table~\ref{tabConst}).
The \textit{mgu} $ \{ \sigma(X_{id}) = X_1 \}$ allows unification between
 $professor(X_1)$ in $body(q_1)$ and $body(c_{P_1})$.
 As    $c_{P_1}$ is triggered and  $\sigma$ is  an homomorphism from $body(c_{P1})$ to $body(q_1)$,
the rewriting step is done by adding $\sigma(head(c_{P_1})) =  teacherOf(X_1,N_1)$ into $body(q_1)$.
Existential variable $Y$ in $head(c_{P_1})$ is replaced by a new fresh variable $N_1$.
This is a step on the chase computation.
Then, the new atom $teacherOf (X_1,N_1)$  triggers constraint $c_{P_2}$, and so on.
The chase process terminates when a fix-point  is reached.
Then the following query \small $q'_1(X_1) \leftarrow$ $professor(X_1), teacherOf (X_1,N_1), offeredCourseAt(N_1,$ $N_2),$ 
			 $employeeGov (X_1),$ $ bornIn(X_1, Z_1), foreignCountry(Z_1)$ \normalsize
 is sent to data providers. 
	Answers for $q'_1$ satisfy both $q_1$ and all the three  above constraints.~$\hfill\Box$	
	
\end{example}


\begin{example}
	\rm
	\label{ex_case3} Let us consider $q_1$ in a new context,  composed of $c_{P_1}$(Table~\ref{tabConst}) and a new $c_{N_2}:$ $ teacherOf(X,Y), headOf(X,Z) \rightarrow \bot .$
	Initially, the rewriting  step is identical to the one discussed in  Example~\ref{ex_case1}:  $ teacherOf(X_1,N_1)$ is added to the query's body. 
	This new atom triggers $c_{N2}$.
To avoid  dealing with negative atoms in the rewritten query,  we replace the negative constraint $c_{N2}$ by a new positive constraint $c_{aux_1}: teacherOf(X,Y) \rightarrow aux\_teacherOf(X,Y)$, where
 \textit{aux\_teacherOf} is an auxiliary relation containing tuples which  respect $c_{N_2}$
 (according to the datasets in Table~\ref{figEx1}, \textit{aux\_teacherOf} contains data about $Peter$, $Tom$ and $Alice$).
We then treat the negative constraint $c_{N2}$ as a positive one.
The rewritten query 
\small $q'_1(X_1) \leftarrow$ $professor(X_1), teacherOf (X_1,N_1), $ $ aux\_teacherOf(X_1,$ $N_1),$ 
			$bornIn(X_1, Z_1), foreignCountry(Z_1)$  \normalsize
is evaluated on $\distFDB$, composed by instances shown in Table~\ref{figEx1},  together with
	the auxiliary database instance $\source_{Aux}$.$\hfill\Box$	
\end{example}

\begin{example}
	\rm
	\label{ex_case2} 
Constraint $c_{P4}$ imposes restrictions on teachers who teach \textit{database}: they should do research in the database domain. No restriction is imposed on teachers in other domains.
For query \small $q_2(X) \leftarrow teacherOf(X,Y)$ \normalsize the situation is different from the one in  Example~\ref{ex_case1}: the unifier between $body(q_2)$ and  $body(c_{P4})$ is not a homomorphism from $body(c_{P4})$ to $body(q_2)$ (but the inverse). 
We cannot apply the reasoning of Example~\ref{ex_case1}: a query such as 
\small $q'_2(X) \leftarrow teacherOf(X,DB),$ $ researchesIn(Z,DB)$ \normalsize would ignore the teachers of all other domains.
Here, we replace $q_2$ by  the two  queries, namely,
\small $q_{2.1}(X) \leftarrow teacherOf(X,Y),$ $Y \neq DB$ \normalsize and 
\small $q_{2.2}(X) \leftarrow teacherOf(X,DB), researchesIn(X,DB).$ \normalsize
	The first query is created by adding a  comparison atom  to deal with results that are \textit{not} concerned by the constraint.  The second query is created  by integrating  $head(c_{P4})$ in $q_2$ and applying a  comparison atom 
	($Y = DB$)  	which covers the results related to the constraint.$\hfill\Box$
\end{example}

In the rest of this section,  we explain the preprocessing of  negative constraints, we show the details of our rewriting method  proving  that it is sound and complete and we present the whole validation process via rewriting.

\subsection{Preprocessing Negative Constraints}
\label{subsection:PreProcessingNC}

The translation of negative constraints implies computing auxiliary instances.

%
%
%
%
%

\begin{definition}[Auxiliary Relations from Negative Constraints]
	\label{def-AuxRelation}
	\linebreak{\rm
	Let $L_1(u_1), L_2(u_2), comp \rightarrow \bot$  be  a negative constraint 
		with $var(u_1) \cap var (u_2) = \{x_1, \dots, x_n\}$
		(we denote a sequence such as $x_1, \dots, x_n$ by a variable in capital letter).	
		Auxiliary relations $\auxR$, and, resp.,  $\auxRI$,  denoted as the  complementary relation of $L_1$ \wrt\ $L_2$ 
		(resp., of  		$L_2$ \wrt\ $L_1$) are obtained by the evaluation of the following datalog programs on a given database source
		$\distFDB$.
		
		\noindent
		\begin{tabular}{lcl}
			Computation of $\auxR$:	& \hspace{1cm} 	& Computation of $\auxRI$:\\
			$q_1(X) \leftarrow L_1(X, Y)$				&&$q_1(X) \leftarrow L_2(X, Y)$\\
			$q_2(X) \leftarrow L_2(X,Y), comp|_{L_2}$	&& $q_2(X) \leftarrow L_1(X,Y), comp|_{L_1}$\\
			$q_3(X) \leftarrow q_1(X), \neg q_2(X)$		&& $q_3(X) \leftarrow q_1(X), \neg q_2(X)$\\
			$\auxR(X, Y) \leftarrow L_1(X, Y), q_3(X)$	&&$\auxRI(X, Y) \leftarrow L_2(X, Y), q_3(X)$

		\end{tabular}
		
		\noindent
		or, equivalently, the evaluation of the following relational algebra queries:\\
		$~~~\auxR := L_1 \Join ( \Pi_X(L_1) \setminus \Pi_X (\sigma_{comp|L_2}(L_2))$ and \\
		$~~~\auxRI := L_2 \Join ( \Pi_X(L_2) \setminus \Pi_X (\sigma_{comp|L_1}(L_1))$ \hfill$\Box$

	}
\end{definition}

Let $c_N$ be a negative constraint. 
Algorithm \textit{Preprocessing Negative Constraints} (detailed in~\cite{BinhTHESE}) deals with different situations.
Firstly, when the body of $c_N$ contains facts, we distinguish two scenarios.
If  these facts do not exist in the database instance 	$\distFDB$, the constraint $c_N$ is not added into the set $C_{Nrw}$. 
Otherwise, if these facts are true in 	$\distFDB$, for each fact $f$ in $body(c_N)$ we create a new constraint of 	the format $f \rightarrow \bot$.
Finally, when the body of $c_N$ is not composed by facts,  new constraints are added to $C_{Nrw}$.
If $c_N$ has only one atom in its body; $c_N$ itself is added to $C_{Nrw}$.
	Otherwise, when $c_N$ has two atoms in its body (\ie, $L_1(u_1), L_2(u_2) \rightarrow \bot$); 	$c_N$ is translated into two new constraints. 
Each new constraint involves an auxiliary complementary relation. 
	Relation $\auxR (u_1)$ ($\auxRI (u_1)$) contains the complement of $L_1$ \wrt\ $L_2$ 
	(respectively, $L_2$ \wrt\ $L_1$ )
	and the join and selection conditions appearing in $c_N$. 
	Definition~\ref{def-AuxRelation} establishes the queries that should be evaluated in order to compute instances of $\auxR (u_1)$ and $\auxRI (u_1)$ (to be stored in $\source_{aux}$). All instances in $\source_{aux}$ are computed on the basis of the current instance $\distFDB$.

\begin{example}
	\rm
	Consider the dataset \distFDB\  
	and the new negative constraint: \\
	\small
	\centerline{$c_{N_3}: worksFor(X, Y, Z), bornIn(X, W), (Z = CNRS), (W=France) \rightarrow \bot. $} 
	\normalsize 
	According to Definition~\ref{def-AuxRelation},  auxiliary relations 
	\small
	$aux\_worksFor^{c_{N_3}} $ 
	\normalsize
	and \small
	$aux\_bornIn^{c_{N_3}}$ 
	\normalsize
	are computed by the rules
	\small
	$aux\_worksFor^{c_{N_3}}(X, Y, Z) \leftarrow worksFor(X, Y, Z), q_{13}(X)$	 \normalsize and 
	\small $aux\_bornIn^{c_{N_3}}(X, W) $ $\leftarrow bornIn(X, W), q_{23}(X)$	  \normalsize with:
	
	\small
	\begin{tabular}{lll}
\hspace{-0.8cm}	$q_{11}(X) \leftarrow worksFor(X, Y, Z);$ &  \hspace{-0.4cm}	$q_{12}(X) \leftarrow bornIn(X, France);$ &  \hspace{-0.5cm}$q_{13}(X) \leftarrow q_{11}(X), \neg q_{12}(X)$\\
\hspace{-0.8cm}		$q_{21}(X) \leftarrow bornIn(X, W);$ &  \hspace{-0.9cm} $q_{22}(X) \leftarrow worksFor(X, Y, CNRS);$	&  \hspace{-0.4cm} $q_{23}(X) \leftarrow q_{21}(X), \neg q_{22}(X).$\\
	\end{tabular}
\normalsize
These relations are added into the auxiliary database $S_{aux}$ and   $c_{N3}$ is transformed into two constraints:
	\small
	$worksFor(X, Y, CNRS) \rightarrow aux\_worksFor^{c_{N3}}(X,Y,Z) $ 
	 \normalsize
	  and 
	  \small
$bornIn(X, France) \rightarrow aux\_bornIn^{c_{N3}}(X,W). $ \hfill $\Box$
	\normalsize
\end{example}

\subsection{Rewriting Query Algorithm}
\label{subsection:AlgoRWC}

Given a query $q$, and a set of constraints $\C_{rw}$, which consists of constraints in $\C_P$ and those resulted from the preprocessing step, 
the rewriting process consists of performing the chase, starting with the atoms in $body(q)$. The idea is similar to the one used in \cite{DPT99}.
Our  algorithm denoted by $RewriteQuery$ (Algorithm~\ref{algoQR}) proposes an iteration over constraints and queries. 
At each iteration, a step of the chase is performed by Algorithm~\ref{algoChSt} to incorporate constraints in $\C_{rw}$ to the body of $q$.

A  constraint $c$ is integrated into  query $q$ if it is triggered by an atom in $q$.
After translating $\C_N$ into $\C_{Nrw}$, constraints have only one atom in their bodies. Thus, testing whether  $c$ is triggered by an atom $L$ in $body(q)$ is to check the existence of $\theta$, a \textit{mgu} which ensures $q$'s satisfiability, between $L(u)$ and $body(c)$ (Algorithm~\ref{algoChSt}, line~\ref{algoChSt:mgu}).
Once the condition on line~\ref{algoChSt:sat} 
is satisfied, the constraint $c$ should be matched to atom $L(u)$. 
To this end, variables in $c$ are renamed by variables in $q$ by using the \textit{one-way mgu} defined below.

\begin{definition}[One-way MGU] \label{specialMgu}
	\rm
	Let $l_1$ and $l_2$ two literals which are unifiable. 
	A one-way unifier denoted by $\theta_{l_1\rightarrow l_2}$ is a $mgu$ for $l_1$ and $l_2$ such that all variables of $\theta_{l_1 \rightarrow l_2}(l1)$ are variables of $l2$. 
	\hfill\(\Box\)
\end{definition}

For example, let $l_1=A(X_1, a, b,Y_1, Y_1)$ and $l_2=A(a,X_2, Y_2, Z_2, U_2)$ which are unifiable. 
A possibility for the one-way mgu from $l_1$ to $l_2$ is $\{ \theta_{l_1 \rightarrow l_2}(X_1) = a, \theta_{l_1 \rightarrow l_2}(Y_1) = Z_2, \theta_{l_1 \rightarrow l_2}(X_2) = a, \theta_{l_1 \rightarrow l_2}(Y_2) = b, \theta_{l_1 \rightarrow l_2}(U_2) = Z_2\}$  
and \linebreak
$\{\theta_{l_2 \rightarrow l_1}(X_1) = a, \theta_{l_2 \rightarrow l_1}(X_2) = a, \theta_{l_2 \rightarrow l_1}(Y_2) = b, \theta_{l_2 \rightarrow l_1}(Z_2) = Y_1, \theta_{l_2 \rightarrow l_1}(U_2) = Y_1\}$  for the one-way mgu from $l_2$ to $l_1$.
Clearly $\theta_{l_1 \rightarrow l_2}(l1) = \theta_{l_1 \rightarrow l_2}(l2)$.\\

The notion of \textit{one-way} unifier is essential when constraints have constants or repeated variables in their bodies. Algorithm~\ref{algoChSt}, line~\ref{algoChSt:homoh}, finds the homomorphism $h$ once the one-way unifier is applied to the body of $c$.
The need of the \textit{one-way mgu} is illustrated by the example below.

\begin{example} 
	\rm
	Let $c: A(X_1, a, b,Y_1, Y_1) \rightarrow B(X_1, Y_1)$ be a constraint 
	and let \small $q(X_2, Y_2, Z_2, U_2) \leftarrow A(a,X_2, Y_2, Z_2, U_2), C(U_2)$ \normalsize be a query.
	
		Atoms $l_1=A(X_1, a, b,Y_1, Y_1)$ and $l_2=A(a,X_2, Y_2, Z_2, U_2)$ are unifiable 
	but there is neither a homomorphism from the $l_1$ (\ie, $c$'s body) to $l_2$
	nor a homomorphism from $l_2$ to $l_1$.
	However, since \textit{mgu} $\theta$ exists, it  is possible to find an instantiation of $A(a,X_2, Y_2, Z_2, U_2)$ that triggers $c$. 
	In other terms, an answer for $q$ can trigger $c$ imposing the constraint verification.
	
	An one-way unifier, such as 
	$\{\theta_{l_1 \rightarrow l_2}(X_1) = a, \theta_{l_1 \rightarrow l_2}(Y_1) = Z_2, \theta_{l_1 \rightarrow l_2}(X_2) = a, \theta_{l_1 \rightarrow l_2}(Y_2) = b, \theta_{l_1 \rightarrow l_2}(U_2) = Z_2\}$ 
	applied to $c$ gives
	$A(a, a, b,Z_2, Z_2) \rightarrow B(a, Z_2)$.
	Now it is possible to find a homomorphism from $l_2$ to $l_1$.

	An one-way unifier allows Algorithm~\ref{algoChSt} to produce new queries $q_1$ and $q_2$, according to lines~\ref{SpecialChase}-\ref{endSpecialChase}. 
	Let us consider the homomorphism $h$ from $l_2$ to $\theta_{l_1 \rightarrow l_2}(l_1)$ such that:
	$h(X_2) = a,$ $h(Y_2) = b$, $h(Z_2) = Z_2$, $h(U_2) = Z_2$.
	Then, according to Algorithm~\ref{algoChSt} 
	we obtain the following queries: 
	
	\small
	\begin{enumerate}
	    \item[ ]\hspace{-0.5cm}$q_2:$ $ q(X_2, Y_2, Z_2, U_2) \leftarrow A(a,X_2, Y_2, Z_2, U_2), C(U_2), B(a, Z_2), 
		X_2=a, Y_2=b, U_2=Z_2 $
		\item[ ]\hspace{-0.5cm}$q_{1_1} :$ $q(X_2, Y_2, Z_2, U_2) \leftarrow A(a,X_2, Y_2, Z_2, U_2), C(U_2), X_2 \neq a$
		\item[ ]\hspace{-0.5cm}$q_{1_2} :$ $ q(X_2, Y_2, Z_2, U_2) \leftarrow A(a,X_2, Y_2, Z_2, U_2), C(U_2), Y_2 \neq b$ 
		\item [ ]\hspace{-0.5cm}$q_{1_3} : $ $q(X_2, Y_2, Z_2, U_2) \leftarrow A(a,X_2, Y_2, Z_2, U_2), C(U_2), U_2 \neq Z_2$ 
	\end{enumerate}
	\normalsize

	The one-way unifier allows rewriting a query on the basis of the unification of only two atoms - one in the query's body,  the other in the constraint's body. Without it, it would be necessary to apply a $mgu$ to the whole query's body at each step.~\hfill\(\Box\)

\end{example}

\begin{algorithm}[]
	\caption{A step of chase processing }
	\label{algoChSt}
	\SetKwInOut{Input}{Input}
	\SetKwInOut{Output}{Output}
	\Input {A conjunctive query $q:\tau_{in}$ and a constraint $c \in \C_{rw}$ where $\C_{rw} = \C_P \cup \C_{Nrw}$ }
	\Output {A set $Q$ of rewritten queries $q_r:\tau_{in}$, such that for each $q_r:\tau_{in}$ we have $head(q_r) = head(q)$.
	}
	\SetKwFunction{StCh}{StepChase}
	\SetKwProg{Fn}{Function}{:}{\KwRet}
	\Fn{\StCh{$q:\tau_{in}$, c}}{
		\ForEach{atom $L(u) \in body(q)$ such that:\\
			there is a mgu $\theta: \theta(L(u))=\theta(body(c))$ for which \label{algoChSt:mgu}
			$\theta(body(q))$ is satisfiable \label{algoChSt:sat}} {
			Denote query $q:\tau_{in}$ by $q(u_0) \leftarrow \beta(u_1), L(u) $\;
			Let $\sigma_c$ be $\theta_{body(c) \rightarrow L(u)}$ (Definition~\ref{specialMgu}) \label{algoChSt:sigmaC} \;
			\If{$ \exists\ homomorphism\ \nu\ from\ body(\sigma_c(c))\ to\ L(u)$ \label{testTrigger} \label{algoChSt:testHomoNu} \label{NormalChase}}{
				\If {$\neg Isomorphic(\nu(\sigma_c((head(c))), q)$ \label{algoChSt:isom}}{
					$q(u_0) \leftarrow \beta(u_1), L(u), (\nu(\sigma_c((head(c)))$ \label{completeQ}\label{endNormalChase}\;
					$ Q:= Q \cup \{q:\tau_{in}\} $\;
				}
			}
			\Else{
				Let $h$ be a homomorphism such that $h(L(u)) = \sigma_c(body(c))$ \label{algoChSt:homoh} \label{SpecialChase}\;
				Let $ Q:= \emptyset$\;
				Let $ q_2:\tau_{in} $ be $q(u_0) \leftarrow \beta(u_1), L(u), h(\sigma_c(head(c)))$\;
				\ForEach {var. $x$ in $u$ for which $h(x) = a$, where $a$ is a term
					 $a \neq x$ \label{algoChSt:foreachq_1}}
				{	Let $q_1:\tau_{in} $ be $ q(u_0) \leftarrow \beta(u_1), L(u) , (x \neq a)$\;
					$ Q:= Q \cup \{q_1:\tau_{in} \} $\;
					$body(q_2) := body(q_2) \wedge (x = a)$\;
				}\label{algoChSt:foreachq_1_end}
				$ Q:= Q \cup \{q_2:\tau_{in} \} $\ \label{endSpecialChase};
			}			
		}
		return($Q$)\;
	}
	
\end{algorithm}

Once $\theta$ and $\theta_{body(c) \rightarrow L(u)}$ are determined on line~\ref{algoChSt:sigmaC} of Algorithm~\ref{algoChSt}, 
we test if there is a homomorphism
$\nu$ from $body(\sigma_c(c))$  to $L(u)$.
This is the criteria to decide \textit{how} $c$  is integrated into $q$ (line~\ref{algoChSt:testHomoNu}). 
Indeed, we have two main situations:

\vspace{0.2cm}
\noindent
(1)
When the homomorphism $\nu$ exists, the query's body is completed with the atom computed from the head of $\sigma_c(c)$ (line~\ref{completeQ}). 
	The new atom is added into $body(q)$ if it is \textit{not}  isomorphic to any atom already existing in $q$ (line~\ref{algoChSt:isom}).
	In this way,  we  avoid   adding redundant atoms into the body of the query, but we allow the result of the rewriting step to be empty
	(since $c$'s consequence may already be in $body(q)$).
 We recall that our constraints respect some syntactic restrictions, namely:
	(A)  $\C_P$ is a set of weakly acyclic TGD \cite{Fagin2005}:
		 it does not allow infinite labelled null creation during the chase, 
(B) rules resulting from the translation of negative constraints		cannot trigger other constraints: their head's predicates are new, and do not belong to the initial  schema. 
	
\vspace{0.2cm}
\noindent
(2) When the homomorphism $\nu$ does not exist,   one of the following situations hold:
	(i) we cannot map  a constant in $body(c)$  to  a variable or a different constant in $L(u)$ or
	(ii) we cannot map  a variable in $body(c)$ (appearing more than once) to different variables in $L(u)$.
	For instance, in   Example~\ref{ex_case2}, 
	no homomorphism from $body(\sigma_c(c_{p2}))$ to $L = teacherOf(X,Y)$ is possible.
	However, as there is a \textit{mgu} between $L$ and $body(c)$, there is  a homomorphism $h$ from $L$ to $body(\sigma_c(c_{p2}))$ (line~\ref{algoChSt:homoh}).
	In this case, we generate two queries, namely:
	(A) a query ($q_2$ in the Algorithm~\ref{algoChSt}) that deals with results (\ie, possible query instantiations)
		involving  constraint $c$, and 
(B) a set of queries ($q_1$ in the \textit{foreach} loop at line~\ref{algoChSt:foreachq_1}-\ref{algoChSt:foreachq_1_end}) dealing  with results that are \textit{not} concerned by $c$.
In both cases,  $c$ can be written by using  comparison  atoms  ($comp$)  containing equalities.
	During the rewriting process, Algorithm~\ref{algoChSt} generates  queries
	$q_1$  by adding into $body(q)$  the negation of these comparison atoms. 
	For instance, in our Example~\ref{ex_case2}, $q_{2.1}$ selects people who do not teach $DB$.
	With the database instance of Table~\ref{figEx1}, the answer for $q_{2.1}$ is $Tom, Alice, Peter$. 
	Query $q_{2.2}$ deals with results \textit{concerned} by the constraint. It selects two kinds of people:
	$(i)$ those who are database researchers and \textit{only} teach $DB$ and
	$(ii)$ those who teach and do research in the database domain but also teach other subjects.
	In this example, the answers for $q_2$ on the instance of Table~\ref{figEx1}
	are $Bob$, $Tom$, $Alice$ and $Peter$.
	Notice that $Bob$ is not an answer for $q_{2.1}$, but it is the answer to $q_{2.2}$. The result of $q_2$ is the union of the answers for $q_{2.1}$ and $q_{2.2}$.
Rewritten queries dealing with results not concerned by c (\ie\ $q_1$ in Algorithm~\ref{algoChSt}) introduce negative $comp$ atoms which, 
in the relational algebra,  can be translated to a  selection with a negative condition.

\vspace{0.2cm}
Let us now consider in detail Algorithm~\ref{algoQR}. 
The general idea here is to apply exhaustively the chase step with constraints in $C_{rw}$ on $q$.
Applying the chase step on $q$ and a constraint $c \in C_{rw}$ results in a set of rewritten queries $Q'$ (line~\ref{algoQR:Q'}) which may replace $q$ in the next step. 
Recall that the purpose of integrating constraint $c$ into $q$ by applying the chase step on $q$ and $c$ is to filter answers of $q$ that are valid with respect to $c$.

\begin{algorithm}[ht]
	\caption{Query Rewriting}	
	\label{algoQR}
	\SetKwInOut{Input}{Input}
	\SetKwInOut{Output}{Output}
	\Input {A conjunctive query $q:\tau_{in}$ and a set of constraints $\C_{rw}= \C_P \cup \C_{Nrw}$ }
	\SetKwFunction{RQ}{RewriteQuery}
	\SetKwProg{Fn}{Function}{:}{\KwRet}
	\SetKwFunction{StCh}{StepChase}
	
	\SetKwProg{Fn}{Function}{:}{\KwRet}
	\Fn{\RQ{$q:\tau_{in}$, \C}}{
		$Q = \{q:\tau_{in}\}$\;
		\Repeat{not Changed}{
			$Changed := false$\;
			\ForEach{ $c \in \C$ \label{chaseStep}}{
				\ForEach{ $q:\tau_{in} \in Q$ } {
					$replace := false$\;
					$Q' :=$ \StCh{$q:\tau_{in}$, c}\; \label{algoQR:Q'}	
					\tcc{Algorithm~\ref{algoChSt}}		
					\ForEach{ $q':\tau_{in} \in Q'$ \label{algoQR:replace} }{
						\If{ ($q':\tau_{in}$ is more restricted than $q:\tau_{in}$) and ($q':\tau_{in}$ is not contradictory) \label{algoQR:condRestricted}}{
							$Q:= Q \cup \{q':\tau_{in}\} $\; 
							$Changed := true$\;
							$replace :=true$\;
						}
					} 
					\If{$replace$} {				
						$Q:= Q \setminus \{q:\tau_{in}\}$ \label{algoQR:replaceEnd} \;	
					}	
				} 
			}
		}
		\KwRet Q\;
	}
\end{algorithm}

Query $q$ is replaced by $Q'$ only if $Q'$ is more restricted than $q$, \ie\ $ans(Q')$ is included in $ans(q)$ (line~\ref{algoQR:condRestricted}).
On the other hand, in Algorithm~\ref{algoChSt}, for a given query $q$, each rewritten query $q'$ is created by instantiating atoms or adding more atoms \ignore{(and/or comparison conditions)} into $body(q)$. 
Thus, an answer to each $q'$ is either equal to or included in the answer to $q$.
When $q'$ is equivalent to $q$, we can remove it from $Q'$. In other words, only rewritten queries (strictly) more restricted than $q$ are considered replacing $q$ in $Q$ (lines~\ref{algoQR:replace}-\ref{algoQR:replaceEnd}).
The new set $Q$, in turn, will be chased with every constraint $c \in C_{Nrw}$, including constraints already considered in previous steps, until 
no  new query is generated. 


\begin{example}
	\rm
	Consider query \small $q(X) \leftarrow L_1(X, Y)$ \normalsize and two constraints $c_1 : L_1(X, Y)$ $ \rightarrow L_2(X, Y)$  and $c_2 : L_2(X, a) \rightarrow L_1(X, b)$.
	Figure~\ref{algoQR:ex} illustrates the execution of Algorithm~\ref{algoQR}, step by step: each column points to the execution of a specific line.
	The sixth column shows $q'$ in $Q'=stepChase(q,c)$ such that $q'$ is satisfiable and more restricted than $q$.
	Intermediate rewritten queries are depicted in Figure~\ref{algoQR:exRWlist}.
Notice that queries $q'_{21}$, $q'_{22}$ and $q'_{31}$ are removed from $Q'$ by the \textit{isomorphic condition} in Algorithm~\ref{algoChSt} (line~\ref{algoChSt:isom}), \ie\ there exists an atom in $body(q)$ that is isomorphic to the added atom in the rewriting step. 
	Queries $q_{41}$ and $q_{42}$ are removed by the condition at line~\ref{algoQR:condRestricted} in Algorithm~\ref{algoQR}.
	Thus, the rewriting result of $q$ \wrt\ $c_1$ and $c_2$ is the set of two following rewritten queries:
	\small $q_{21}(X) \leftarrow L_1(X,Y), L_2(X,Y), Y \neq a$ \normalsize and \\ 
	\small $q_{31}(X) \leftarrow L_1(X,a), L_1(X,b), L_2(X,a), L_2(X,b)$\normalsize. $\hfill\Box$

\begin{figure}
\footnotesize
\centering
\begin{tabular}{c}
\subfloat[	Intermediate results of Algorithm~\ref{algoQR}]
		{\scalebox{0.7}{	
		\begin{tabular}{|c|c|l|l|l|l|l|l|}\hline
		Line3 & Line5& $Q$(line3)& $q \in Q$ & $Q'$(line8) & $q'$(line10)& $Q$(line16)&\textit{Changed}\\ \hline
			1	& $c_1$	& \{$q$\}		& $q$	 & \{$q_1$\}	 & \{$q_1$\}	& \{$q_1$\}	 & T	\\ \hline
			& $c_2$	& \{$q_1$\}		& $q_1$	& \{$q_{21}$,$q_{22}$\}	& \{$q_{21}$,$q_{22}$\}	 & \{$q_{21}$,$q_{22}$\}	 & T	\\ \hline
			2	& $c_1$	& \{$q_{21}$,$q_{22}$\} 		& $q_{21}$	 & \{ \}	& \{ \}	& \{$q_{21}$,$q_{22}$\}	 & F	\\ 
			&		&			& $q_{22}$	 & \{$q_{31}$\}	& \{$q_{31}$\}	 & \{$q_{21}$,$q_{31}$\}	 & T	\\ \hline
			& $c_2$	& \{$q_{21}$,$q_{31}$\} & $q_{21}$	 & \{$q_{41}$,$q_{42}$\}	& \{ \}	& \{$q_{21}$,$q_{31}$\}	 & F	\\ 
			&		&			& $q_{31}$	 & \{ \}	& \{ \}	& \{$q_{21}$,$q_{31}$\}	 & F	\\ \hline
		\end{tabular}
		}
		\label{algoQR:ex}
		} \\
\subfloat[	Intermediate rewritten queries]	
	{\scalebox{0.8}{
		\begin{tabular}{|l|l|l|}
			\hline
			Algo.~\ref{algoChSt} & $q'$ & Remark \\ \hline
			& $q(X) \leftarrow L_1(X,Y)$ &  \\ \hline
			$(q,c_1)$ & $q_1(X) \leftarrow L_1(X,Y), L_2(X,Y)$ &  \\ \hline
			$(q_1,c_2)$ & $q_{21}(X) \leftarrow L_1(X,Y), L_2(X,Y), Y \neq a$ &  \\
			& $q_{22}(X) \leftarrow L_1(X,a), L_2(X,a), L_1(X,b)$ &  \\ \hline
			$(q_{21},c_1)$ & $q'_{21}(X) \leftarrow L_1(X,Y),L_2(X,Y),L_2(X,Y)$ & isomophism \\ \hline
			$(q_{22},c_1)$ & $q'_{22}(X) \leftarrow L_1(X,a), L_2(X,a), L_2(X,a), L_1(X,b)$ & isomophism \\ \hline
			$(q_{22},c_1)$ & $q_{31}(X) \leftarrow L_1(X,a), L_2(X,a), L_1(X,b), L_2(X,b)$ &  \\ \hline
			$(q_{21},c_2)$ & $q_{41}(X) \leftarrow L_1(X,Y), L_2(X,Y), Y \neq a, Y\neq a$ & \begin{tabular}[c]{@{}l@{}}not more \\ restricted than $q_{21}$\end{tabular} \\
			& $q_{42}(X) \leftarrow L_1(X,Y), L_2(X,Y), Y=a, L_1(X,b), Y \neq a$ & contradictory \\ \hline
			$(q_{31},c_2)$ & $q'_{31}(X) \leftarrow L_1(X,a), L_2(X,a), L_1(X,b), L_1(X,b), L_2(X,b)$ & isomophism \\ \hline
		\end{tabular}
		}			
		\label{algoQR:exRWlist}
		}\\
\end{tabular}
\caption{Running Algorithm~\ref{algoQR}.}
\vspace*{-0.5cm}
\end{figure}

\end{example}


\subsection{Correctness of Validation through rewriting queries}
\label{subsection:Proof}
The complete proof of the correctness and completeness of Algorithm~\ref{algoQR} is presented in~\cite{BinhTHESE} (page 76).
It concerns the following two propositions.

\begin{proposition}
	{\rm
		Given a conjunctive query $q$ and a set of constraints $\C_{rw}$, 
		Algorithm~\ref{algoQR} finishes in a finite number of steps. $\hfill\Box$
	}
\end{proposition}

\noindent
\textsc{proof :} The rewriting algorithm is composed of two parts.
The first part deals with  situations where the homomorphism $\nu$ exists. Therefore, when
our rewriting algorithm is  restricted to  these situations
(\ie, when only lines~\ref{NormalChase} to~\ref{endNormalChase} are executed in
function \textsf{StepChase} (Algorithm~\ref{algoChSt})),  it corresponds to the chase with non weakly acyclic constraints $\C_1$ (tgd)  of a conjunctive query $q$ (and therefore to the chase phase of algorithms introduced in~\cite{Deutsch06}).
Thus, as in~\cite{Deutsch06},  which is based on proofs in~\cite{Fagin2005},  it terminates.
The second part  considers the situation where the homomorphism $\nu$ does not exist.
In this situation, a query $q$ generates at least two other queries. Clearly,  the number of new queries that can be created during the execution of lines~\ref{SpecialChase} to~\ref{endSpecialChase} is finite and bounded by the number of  variables for which conditions on line~\ref{algoChSt:foreachq_1} hold.
From these observations and since the number of constraints and queries is finite, Algorithm~\ref{algoQR} terminates. Moreover, after each execution of Algorithm~\ref{algoChSt} the new generated queries are compared to those already existing, allowing to keep only the most restricted 
ones (along the lines of~\cite{AHV95}).~\hfill$\Box$


\begin{proposition}
	\label{prop:RewCorrectComplete}
	{\rm
		Let $\C_1$ be the set of constraints obtained from a given context \C\ (where $\C = \C_P \cup \C_N$) by the preprocessing of negative constraints.
		Let $\distFDB$ and $S_{aux}$ be, respectively,  the database instance and the auxiliary database  obtained on the basis of Definition~\ref{def-AuxRelation}.
				For any database instance $\distFDB$,  we have 
		$q(\distFDB,\C) = q_r(\distFDB \cup S_{aux})$
		where $q(\distFDB,\C)$ is the set of \textit{valid} answers of $q$ (with respect to \C) on $\distFDB$
		and  $q_r(\distFDB\cup S_{aux})$ is the set of answers  of $q_r$ on $\distFDB \cup S_{aux}$ with respect to the translated set of constraints $\C_1$. 
		\hfill$\Box$
	}
\end{proposition}

\noindent
\textsc{proof:} (Sketch) The proof is done by contradiction and is based on two lemmas.
The first lemma deals with the situation where \C\ has only a positive constraint whose body cannot be mapped to an atom in the query's body (although a unifier exits) (as on line~\ref{algoChSt:homoh} Algorithm~\ref{algoChSt}). We prove that the  corresponding rewritten query $q_r$ computes all and only the $q$'s answers which are valid with respect to \C.
The second lemma proves that when \C\ has only one negative constraint of the form $c: A(v_1), B(v_2) \rightarrow \bot$
whose body is triggered by an atom in the query's body, the corresponding rewritten query $q_r$ computes all and only the $q$'s answers which are valid with respect to \C. 
Moreover, it proves that, in this situation, query $q_r$ involves auxiliary instances computed according to Definition~\ref{def-AuxRelation} and stored in $S_{aux}$. ~\hfill$\Box$

\subsection{The whole validation process}
\label{subsection:wholeProcessing}

The query rewriting process (Algorithms~\ref{algoChSt} and~\ref{algoQR}) only deals with $C_P$ and $C_N$ (besides $\tau_{in}$).  To verify if the results obtained from evaluations of rewritten queries respect constraints in $C_K$, the naive approach is used.
This overall process is depicted in Algorithm~\ref{algoCompleteVal}.

\vspace*{-0.2cm}
\begin{algorithm}[]	
	\caption{ Valid candidate answers }
	\label{algoCompleteVal}
	\SetKwInOut{Input}{Input}
	\SetKwInOut{Output}{Output}
	\Input {$\bullet$ A conjunctive query $ q:\tau_{in} $ and a set of constraints $ \C = C_P \cup C_N \cup C_K $.\\
		$\bullet$ An access to the database instance $\distFDB$}
	\Output {Answers of $q:\tau_{in}$ respecting $\C$.}
	\SetKwFunction{vCA}{valCandAns}
	\SetKwProg{Fn}{Function}{:}{\KwRet}
	\Fn{\vCA{$q:\tau_{in},\C$}}{
	
	    $\textsf{AnsSet} = \emptyset$; \\
	$C_{rw} = C_P \cup RewritingNegConstraints(C_N); \label{algoComVal:resultAlgoRNC}$\\
	$Q = RewriteQuery(q:\tau_{in}, C_{rw})$; \label{algoComVal:resultAlgoRQ}\\
	$\textsf{Solutions} = Eval (Q, \distFDB)$; \label{eval}\\
	$\cache = CreateCache()$; \label{cache}\\
	\ForEach{$ sol \in \textsf{Solutions} $ where $sol = (t, h_t)$}{ \label{subq}
		\If {$\checking (sol,\C_K, \cache, \tau_{in})$ \label{Check}} {
			$\textsf{AnsSet} := \textsf{AnsSet} \cup \{t\}$;
		}
	}
	\KwRet{$ AnsSet$};
}	
\end{algorithm}

\vspace*{-0.2cm}
Function $RewriteQuery$ on line~\ref{algoComVal:resultAlgoRQ} of the Algorithm~\ref{algoCompleteVal} uses constraints in $C_{rw}$ as rewriting rules to reformulate $q:\tau_{in}$ into a set $Q$ of rewritten queries.
Function $Eval$ (line~\ref{eval}) then computes the union of the candidate answers sets for each rewritten query $q':\tau_{in} \in Q$.
These candidate answers are stored in the set $\textsf{Solutions}$.
Note that each candidate answer is computed on $\distFDB$ \wrt\  $C_P$, $C_N$, $\tau_{in}$, and they are stored in $\textsf{Solutions}$ in the form of a pair $(t, h_t)$, where $h_t$ is the homomorphism used to produce tuple $t$ as an answer to a query $q \in Q$.
We need $h_t$ for creating the auxiliary query that is used to validate $t$ in the next step.
Function $Valid$ verifies whether a candidate answer $sol$ is valid \wrt\ $\C_K$ and $\tau_{in}$ on $\distFDB$ by generating corresponding auxiliary queries.
$Valid$ implements the verification concerning $\C_K$ and is performed by the naive algorithm. 
A cache can be used to store the results of auxiliary queries to avoid overcharging data sources (lines~\ref{cache},\ref{Check}).\\


\section{Experimental Study}
\label{chapter:Experiments}

To the best of our knowledge, there is no similar system that supports semantic web data querying and that verifies results to obtain user's quality-driven answers.
Ideas in~\cite{Deutsch06,Fagin2005,Ileana14} are similar to ours, but they are used to other purposes.
Those in~\cite{CGL12,Cali2003,Cali2010,Lembo20153} lack the use of  constraints to filter answers -- our  main contribution.
Comparing our proposal to these systems would require too heavy adaptations.
Hence, our option is to  evaluate  the performance of the different versions of our
\textit{Validator}, namely the \textit{Naive} (mentioned in Section~\ref{section:ValidAnswerFormalDefinition}) and  the \textit{Rewriting} approaches.
Both prototypes have been implemented in Java; developed on the basis of Graal's framework\footnote{https://graphik-team.github.io/graal/ \label{graalRef}}
and tests have been carried out on an HP ZBook with a quad-core Intel i7-4800MQ processors at 2.7GHz with 32KBx4 L1 Cache, 256KBx4 L2 Cache, 6MB L3 Cache, 16GB 799MHz RAM, and a 120 GB hard drive. A 64-bit Windows 10 operating system and the 64-bit Java VM 1.8.031 constitute the software environment.

Experiments are built on the LUBM benchmark\footnote{Lehigh University: http://swat.cse.lehigh.edu/projects/lubm/}, which describes the  structure of universities with 43 classes and 32 properties, and provides a generator of synthetic data with varying size. 
We devised 7 queries and 13  constraints over the LUBM ontology  (5 positive, 5 negative, and 3 key) c.f.\cite{BinhTHESE} (page 89).
Our queries spread from simple ones,  with few atoms ($Q1$, $Q2$) to more complex ones ($Q6$, $Q7$).
The number of  answers vary: $Q2$, $Q4$, $Q6$ give large result sets  while $Q5$, $Q7$ return small ones.
The constraint triggering chain can  involve more than one constraint. Constants appear in some queries or constraints
allowing us to also evaluate the most original aspect of our work: constraints with constants not appearing in queries. 
This is a fine-grained contribution of practical significance -- constants are commonly used in real life scenarios.

With the UBA Data Generator provided by the LUBM, we produced two ABoxesdatasets of different sizes: 
\textit{Dataset 1} contains data of one university, with 86,165 triples and 
\textit{Dataset 5} concerns 5 universities with 515,064 triples.
Produced by LUBM's generator, those datasets are consistent with the inference rules of LUBM, but may be inconsistent with 13 constraints of our experiment.

\begin{table}[ht]
	\centering
	\scalebox{0.56}{
		\begin{tabular}{ccc|
				>{\columncolor[HTML]{C6E0B4}}c |
				>{\columncolor[HTML]{C6E0B4}}c |
				>{\columncolor[HTML]{C6E0B4}}c |
				>{\columncolor[HTML]{C6E0B4}}c |
				>{\columncolor[HTML]{C6E0B4}}c |
				>{\columncolor[HTML]{C6E0B4}}c |
				>{\columncolor[HTML]{F8CBAD}}c |
				>{\columncolor[HTML]{F8CBAD}}c |
				>{\columncolor[HTML]{F8CBAD}}c |
				>{\columncolor[HTML]{F8CBAD}}c |
				>{\columncolor[HTML]{F8CBAD}}c |
				>{\columncolor[HTML]{F8CBAD}}c |}

			\cline{4-9}
			&  &  & \multicolumn{6}{c|}{\cellcolor[HTML]{C6E0B4}\textbf{Dataset 1}} &
			\multicolumn{6}{c|}{\cellcolor[HTML]{F8CBAD}\textbf{Dataset 5}} \\ \cline{2-9} 
			\multicolumn{1}{c|}{} & \multicolumn{1}{c|}{\cellcolor[HTML]{EDEDED}\textbf{\begin{tabular}[c]{@{}c@{}}Num.of \\ atoms in \\ original \\ query\end{tabular}}} & \cellcolor[HTML]{DDEBF7}\textbf{\begin{tabular}[c]{@{}c@{}}Num.of \\ involved \\ constr.\end{tabular}} & \textbf{\begin{tabular}[c]{@{}c@{}}Num.of \\ ans. of \\ original \\ query\end{tabular}} & \textbf{\begin{tabular}[c]{@{}c@{}}Num. \\of \\ valid \\ ans.\end{tabular}} & \textbf{\begin{tabular}[c]{@{}c@{}}Num.of \\ eval. \\ queries\end{tabular}} & \textbf{\begin{tabular}[c]{@{}c@{}}Eval. \\ time\end{tabular}} & \textbf{\begin{tabular}[c]{@{}c@{}}Verif. \\ time\end{tabular}} & \textbf{\begin{tabular}[c]{@{}c@{}}Total \\ time\end{tabular}} 
			&  \textbf{\begin{tabular}[c]{@{}c@{}}Num.of \\ ans.s of \\ original \\ query\end{tabular}} & \textbf{\begin{tabular}[c]{@{}c@{}}Num \\of \\ valid \\ ans.\end{tabular}} & \textbf{\begin{tabular}[c]{@{}c@{}}Num.of \\ eval. \\ queries\end{tabular}} & \textbf{\begin{tabular}[c]{@{}c@{}}Eval. \\ time\end{tabular}} & \textbf{\begin{tabular}[c]{@{}c@{}}Verif. \\ time\end{tabular}} & \textbf{\begin{tabular}[c]{@{}c@{}}Total \\ time\end{tabular}}

			\\ \hline
			\multicolumn{1}{|c|}{Q1} & \multicolumn{1}{c|}{\cellcolor[HTML]{EDEDED}1} & \cellcolor[HTML]{DDEBF7}8 & 1544 & 523 & 6449 & 0.368 & 4.306 & 4.674 & 10095 & 3319 & 42583 & 2.971 & 77.479 & 80.45 \\ \hline
			\multicolumn{1}{|c|}{Q2} & \multicolumn{1}{c|}{\cellcolor[HTML]{EDEDED}1} & \cellcolor[HTML]{DDEBF7}1 & 7861 & 7861 & 7862 & 0.109 & 47.769 & 47.878 & 36682 & 36682 & 36683 & 0.454 & 659.937 & 660.391 \\ \hline
			\multicolumn{1}{|c|}{Q3} & \multicolumn{1}{c|}{\cellcolor[HTML]{EDEDED}2} & \cellcolor[HTML]{DDEBF7}2 & 1822 & 1821 & 3600 & 0.039 & 1.911 & 1.95 & 11900 & 11900 & 23750 & 5.96 & 12.177 & 18.137 \\ \hline
			\multicolumn{1}{|c|}{Q4} & \multicolumn{1}{c|}{\cellcolor[HTML]{EDEDED}2} & \cellcolor[HTML]{DDEBF7}0 & 5939 & 5939 & 1 & 0.033 & 0.364 & 0.397 & 37854 & 37854 & 1 & 0.55 & 2.008 & 5.021 \\ \hline
			\multicolumn{1}{|c|}{Q5} & \multicolumn{1}{c|}{\cellcolor[HTML]{EDEDED}3} & \cellcolor[HTML]{DDEBF7}10 & 50 & 50 & 601 & 0.278 & 0.609 & 0.887 & 59 & 59 & 886 & 1.975 & 2.73 & 4.705 \\ \hline
			\multicolumn{1}{|c|}{Q6} & \multicolumn{1}{c|}{\cellcolor[HTML]{EDEDED}4} & \cellcolor[HTML]{DDEBF7}10 & 6564 & 6564 & 137977 & 8.218 & 127.19 & 135.408 & 36008 & N/A & N/A & 12.617 & oom.err & N/A \\ \hline
			\multicolumn{1}{|c|}{Q7} & \multicolumn{1}{c|}{\cellcolor[HTML]{EDEDED}4} & \cellcolor[HTML]{DDEBF7}9 & 94 & 21 & 502 & 14.512 & 0.296 & 14.808 & 644 & 219 & 4328 & 326.689 & 5.244 & 331.933 \\ \hline
		\end{tabular}
	}
    \\
	\hfill(time in seconds)
	\vspace*{-0.2cm}
	\caption{Evaluation and Verification in the Naive Approach}
	\label{table_naive15univ}
\end{table}

Table~\ref{table_naive15univ} reports the result of the \textit{Naive} approach which consists of two main steps: 
query evaluation to obtain initial results which are, then, checked  \wrt\ constraints,  through sub-queries.
The first two columns indicate, for each query, the number of atoms (\ie,  its complexity) and the number of constraints it triggers.
The remaining columns
 include information about  the execution time and the number of returned results for the original query,   the number of answers which are valid \wrt\  the constraints,
 the number of auxiliary sub-queries needed to perform the constraint verification  and the total time required for the process.

Naive algorithm can handle well small data sets, but,  large datasets may provoke  memory overflow (\eg $Q6$) due to the  great number of auxiliary sub-queries,
generated in a process which is similar to the chase.
Most of our tests show that  the verification step is the most time-consuming step.
For $Q7$,  however, the evaluating time of the complex original query is significantly greater than the validating time  for the small number of results (94 and 644 answers in $Dataset\ 1$ and $Dataset\ 5$, respectively).

\begin{table}[ht]
	\centering
	\scalebox{0.62}{
		\begin{tabular}{ccccc|
				>{\columncolor[HTML]{C6E0B4}}c |
				>{\columncolor[HTML]{C6E0B4}}c |
				>{\columncolor[HTML]{C6E0B4}}c |
				>{\columncolor[HTML]{C6E0B4}}c |
				>{\columncolor[HTML]{F8CBAD}}c |
				>{\columncolor[HTML]{F8CBAD}}c |
				>{\columncolor[HTML]{F8CBAD}}c |
				>{\columncolor[HTML]{F8CBAD}}c |}
			
			&  &  &  & &\multicolumn{4}{c|}{\cellcolor[HTML]{C6E0B4}\textbf{Dataset 1}} &
			\multicolumn{4}{c|}{\cellcolor[HTML]{F8CBAD}\textbf{Dataset 5}} \\ 
			\multicolumn{1}{c|}{} & \multicolumn{1}{c|}{\cellcolor[HTML]{DDEBF7}\textbf{\begin{tabular}[c]{@{}c@{}}Num. \\ of \\ invol. \\ const.\end{tabular}}} &\multicolumn{1}{c|}{\cellcolor[HTML]{DDEBF7}\textbf{\begin{tabular}[c]{@{}c@{}}Num. \\ of \\ rew. \\ queries\end{tabular}}} & \multicolumn{1}{c|}{\cellcolor[HTML]{DDEBF7}\textbf{\begin{tabular}[c]{@{}c@{}}Max num \\ of atomes \\ in rew. \\queries\end{tabular}}} & 
			\multicolumn{1}{c|}{\cellcolor[HTML]{DDEBF7}\textbf{\begin{tabular}[c]{@{}c@{}} Rewriting \\ time \end{tabular}}} & 
			\textbf{\begin{tabular}[c]{@{}c@{}}Num. \\of \\ valid \\ ans.\end{tabular}} &  \textbf{\begin{tabular}[c]{@{}c@{}}Eval. \\ time\end{tabular}} & \textbf{\begin{tabular}[c]{@{}c@{}}Verif. \\ time \\ (for CKs)\end{tabular}} & \textbf{\begin{tabular}[c]{@{}c@{}}Total \\ time\end{tabular}} &
			\textbf{\begin{tabular}[c]{@{}c@{}}Num. \\of \\ valid \\ ans.\end{tabular}} &  \textbf{\begin{tabular}[c]{@{}c@{}}Eval. \\ time\end{tabular}} & \textbf{\begin{tabular}[c]{@{}c@{}}Verif. \\ time \\ (for CKs)\end{tabular}} & \textbf{\begin{tabular}[c]{@{}c@{}}Total \\ time\end{tabular}}			
			
			\\ \hline
			\multicolumn{1}{|c|}{Q1} & \multicolumn{1}{c|}{\cellcolor[HTML]{DDEBF7}8} & \multicolumn{1}{c|}{\cellcolor[HTML]{DDEBF7}1} & \multicolumn{1}{c|}{\cellcolor[HTML]{DDEBF7}8} & \cellcolor[HTML]{DDEBF7}0.017 & 523 &0.473 & 2.094 & 2.584 & 3319 & 1.791 & 10.111& 11.919 \\ \hline
			\multicolumn{1}{|c|}{Q2} & \multicolumn{1}{c|}{\cellcolor[HTML]{DDEBF7}1} & \multicolumn{1}{c|}{\cellcolor[HTML]{DDEBF7}1} & \multicolumn{1}{c|}{\cellcolor[HTML]{DDEBF7}2} & \cellcolor[HTML]{DDEBF7}0.001 & 7861 &0.14 & 0.07 & 0.211 & 36682 & 1.221 & 0.407 & 1.629 \\ \hline
			\multicolumn{1}{|c|}{Q3} & \multicolumn{1}{c|}{\cellcolor[HTML]{DDEBF7}2} & \multicolumn{1}{c|}{\cellcolor[HTML]{DDEBF7}2} & \multicolumn{1}{c|}{\cellcolor[HTML]{DDEBF7}5} & \cellcolor[HTML]{DDEBF7}0.001 & 1821 & 0.109 & 0.078 & 0.188 & 11900 & 1.221 & 0.297 & 1.519 \\ \hline
			\multicolumn{1}{|c|}{Q4} & \multicolumn{1}{c|}{\cellcolor[HTML]{DDEBF7}0} & \multicolumn{1}{c|}{\cellcolor[HTML]{DDEBF7}1} & \multicolumn{1}{c|}{\cellcolor[HTML]{DDEBF7}2} & \cellcolor[HTML]{DDEBF7}0 & 5939 & 0.031 & 0.061 & 0.092 & 37854 & 0.544 & 0.447 & 1.427 \\ \hline
			\multicolumn{1}{|c|}{Q5} & \multicolumn{1}{c|}{\cellcolor[HTML]{DDEBF7}11} & \multicolumn{1}{c|}{\cellcolor[HTML]{DDEBF7}1} & \multicolumn{1}{c|}{\cellcolor[HTML]{DDEBF7}12} & \cellcolor[HTML]{DDEBF7}0.01 & 50 &6.83 & 0.153 & 6.993 & 59 & 9.37 & 0.249 & 9.629 \\ \hline
			\multicolumn{1}{|c|}{Q6} & \multicolumn{1}{c|}{\cellcolor[HTML]{DDEBF7}14} & \multicolumn{1}{c|}{\cellcolor[HTML]{DDEBF7}1} & \multicolumn{1}{c|}{\cellcolor[HTML]{DDEBF7}12} & \cellcolor[HTML]{DDEBF7}0.005 & 6564 & 25.296 & 19.351 & 44.652 & 35922 & 85.293 & 113.451 & 198.749 \\ \hline
			\multicolumn{1}{|c|}{Q7} & \multicolumn{1}{c|}{\cellcolor[HTML]{DDEBF7}11} & \multicolumn{1}{c|}{\cellcolor[HTML]{DDEBF7}1} & \multicolumn{1}{c|}{\cellcolor[HTML]{DDEBF7}12} & \cellcolor[HTML]{DDEBF7}0.008 & 21 & 1.427 & 0.313 & 0.992 & 219 & 14.561 & 1.417 & 15.986 \\ \hline
		\end{tabular}
	}
	\\
	\hfill(time in seconds)
	\caption{Evaluation and Verification of Rewriting Approach}
	\label{table_rewrite1_5univ}
\end{table}

We now turn our attention to the results of  the rewriting approach which consists of four steps.

\begin{itemize}
	\item The \textit{preprocessing step} is performed once for each user-setting-context on a database instance. 
	Auxiliary relations either are materialized and managed in the local system or stored as views computed at evaluating-time. 
	To simplify, the former was realized in the current implementation. Naturally, the preprocessing time is directly proportional to the size of the dataset. 
	Clearly, the number of constraints may increase after this step. 

	\item The \textit{rewriting step} uses positive constraints and transformed negative constraints as rewriting rules to reformulate the initial query.
	This step is completely independent of data in sources. 
	\ignore{The results of this step are reported in Table~\ref{table_rewrite1_5univ}, which contains the following information: 
	(i) the number of involved constraints of the query (note that they may be different from those of the naive case), 
	(ii) the number of the rewritten queries in the result set of this step, and 
	(iii) the largest number of atoms in a rewritten query, which demonstrates that the more constraints are used in the rewriting procedure, the more complex are the rewritten queries  (number of atoms or joins),
	(iv) the time need for rewriting. }
	Theoretically, in the worst case, the number of reformulations of a query can grow exponentially, because each constant in a constraint can lead to two new reformulations. 
	However, thanks to the test conditions for containment and contradiction in \textit{Query Rewriting Algorithm} (line~\ref{algoQR:condRestricted} Algorithm~\ref{algoQR}), only useful reformulations are accepted. 
	Our experimental results proved that the number of rewritten queries can be very small even if the number of involved constraints is not small ($Q1$, $Q5$, $Q6$, $Q7$ in Table~\ref{table_rewrite1_5univ}).	
	By the way, although rewritings are very fast and, in all cases,   the total time is mostly impacted by the evaluation and verification time, 	the rewritten-query complexity affects the evaluation time (\eg evaluation time for the rewritten queries of Q6 and Q7, with 12 atoms in their bodies is high). 

	\item The two last steps are \textit{the evaluation} of rewritten queries followed by \textit{the verification} of the obtained results \wrt\ key constraints. 
	The first noteworthy result is that the rewriting approach has produced answers in all tested cases.
	Moreover, as expected, as in the naive approach, the total processing time is directly proportional to the size of the dataset. 
\end{itemize}

	

\begin{figure}[h]
	\centering
	\includegraphics[width=1\textwidth]{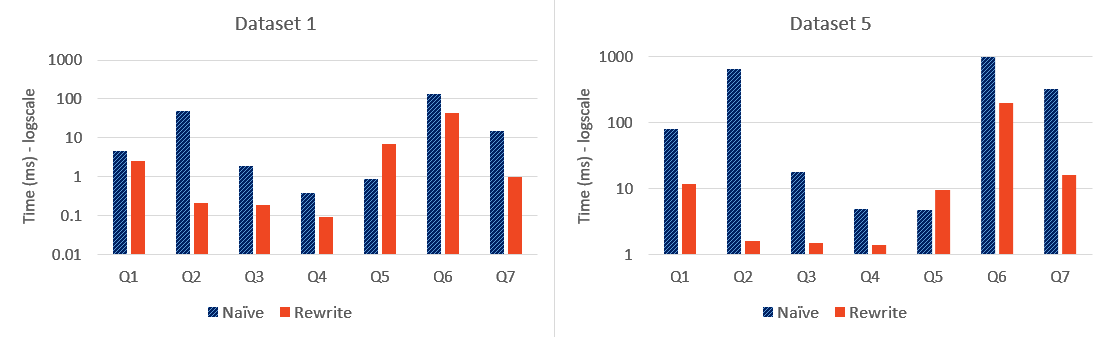}
	\caption{Comparison of total time of Naive approach and Rewrite approach}
	\label{figCompareTotalTime}	
	\vspace*{-0.4cm}
\end{figure}

Figure~\ref{figCompareTotalTime} is a summary of the total processing time for both approaches. 
Whether working on small or large dataset, whether treating simple (\eg\ $Q1$, $Q2$) or complex (\eg\ $Q6$, $Q7$) initial query, whether having few (\eg\ $Q2$, $Q4$) or many (\eg\ $Q6$, $Q7$) involved constraints, the rewriting approach always requires less processing time. 
 Thus, the rewriting approach is  more efficient than the naive one.  
 This better performance is partially explained by our fined-grained method which treats the presence of  constants by adding atoms (\eg\ $Q7$), a solution
  which can highly reduce querying space.
Naive approach is better only for  $Q5$,  a special case:  $Q5$'s answer set is very small (only 50 answers on the dataset 1, or 59 on the dataset 5).

\ignore{We also evaluate the practicality of our context-driven querying system when dealing with large unequal-trust datasets on a distributed environment using MapReduce with HiveQL see \cite{BinhTHESE} for more details.}

\section{Conclusions}
\label{sec:conclusions}
This paper details our solution for validating a set of user quality constraints and confidence degrees when
performing query evaluation. We present it formally and discuss experimental results demonstrating that the 
validation cost is reasonable, and further reduced with our proposed rewriting techniques.  
This is true in particular when constants appear in queries and constraints, and even for constraints with 
constants not appearing in queries. This is the most original aspect of our work, of practical significance as 
constants are commonly used in real life scenarios. As a future work we plan plan to study to which extent our solution would apply to the verification of some SHACL constraints on results from 
SPARQL Endpoints.



%
%

\bibliographystyle{llncs2e/splncs04}      
\bibliography{binh-biblio}   


\end{document}